\documentclass[sigconf]{acmart}
\usepackage[acronym]{glossaries}
\usepackage{graphicx}
\usepackage{subcaption}
\usepackage{wrapfig}
\usepackage{glossaries}
\usepackage{pifont}
\usepackage{makecell}
\usepackage{balance}
\usepackage{url}
\usepackage{hyperref}
\makeglossaries

\AtBeginDocument{%
  }

\setcopyright{None}
\copyrightyear{2025}
\acmYear{2025}
\setcopyright{rightsretained}
\acmConference[UIST Adjunct '25]{The 38th Annual ACM Symposium on User Interface Software and Technology}{September 28-October 1, 2025}{Busan, Republic of Korea}
\acmBooktitle{The 38th Annual ACM Symposium on User Interface Software and Technology (UIST Adjunct '25), September 28-October 1, 2025, Busan, Republic of Korea}\acmDOI{10.1145/3746058.3758385}
\acmISBN{979-8-4007-2036-9/2025/09}

\begin{document}

\title[]{Increasing Interaction Fidelity: Training Routines for Biomechanical Models in HCI}

\author{Michał Patryk Miazga}
\email{miazga@uni-leipzig.de}
\orcid{0009-0003-5579-3036}
\affiliation{%
  \institution{ScaDS.AI, Leipzig University}
  \city{Leipzig}
  \country{Germany}}

\author{Patrick Ebel}
\email{ebel@uni-leipzig.de}
\orcid{0000-0002-4437-2821}
\affiliation{%
  \institution{ScaDS.AI, Leipzig University}
  \city{Leipzig}
  \country{Germany}}

\renewcommand{\shortauthors}{Miazga et al.}


\newacronym{CR}{CR}{Computational Rationality}
\newacronym{HCI}{HCI}{Human-Computer Interaction}
\newacronym{GOMS}{GOMS}{Goals, Operators, Methods, and Selection rules}
\newacronym{KLM}{KLM}{Keystroke-Level Model}
\newacronym{MDP}{MDP}{Markov Decision Process}
\newacronym{POMDP}{POMDP}{Partially Observable Markov Decision Process}
\newacronym{RL}{RL}{Reinforcement Learning}
\newacronym{UCD}{UCD}{User-Centered Design}
\newacronym{UI}{UI}{User Interface}
\newacronym{UX}{UX}{User Experience}
\newacronym{UXD}{UXD}{User Experience Design}
\newacronym{XAI}{XAI}{Explainable AI}
\begin{abstract}
Biomechanical forward simulation holds great potential for HCI, enabling the generation of human-like movements in interactive tasks. However, training biomechanical models with reinforcement learning is challenging, particularly for precise and dexterous movements like those required for touchscreen interactions on mobile devices. Current approaches are limited in their interaction fidelity, require restricting the underlying biomechanical model to reduce complexity, and do not generalize well. In this work, we propose practical improvements to training routines that reduce training time, increase interaction fidelity beyond existing methods, and enable the use of more complex biomechanical models. Using a touchscreen pointing task, we demonstrate that curriculum learning, action masking, more complex network configurations, and simple adjustments to the simulation environment can significantly improve the agent’s ability to learn accurate touch behavior. Our work provides HCI researchers with practical tips and training routines for developing better biomechanical models of human-like interaction fidelity.
\end{abstract}





\begin{teaserfigure}
\centering
  \includegraphics[width=\textwidth]{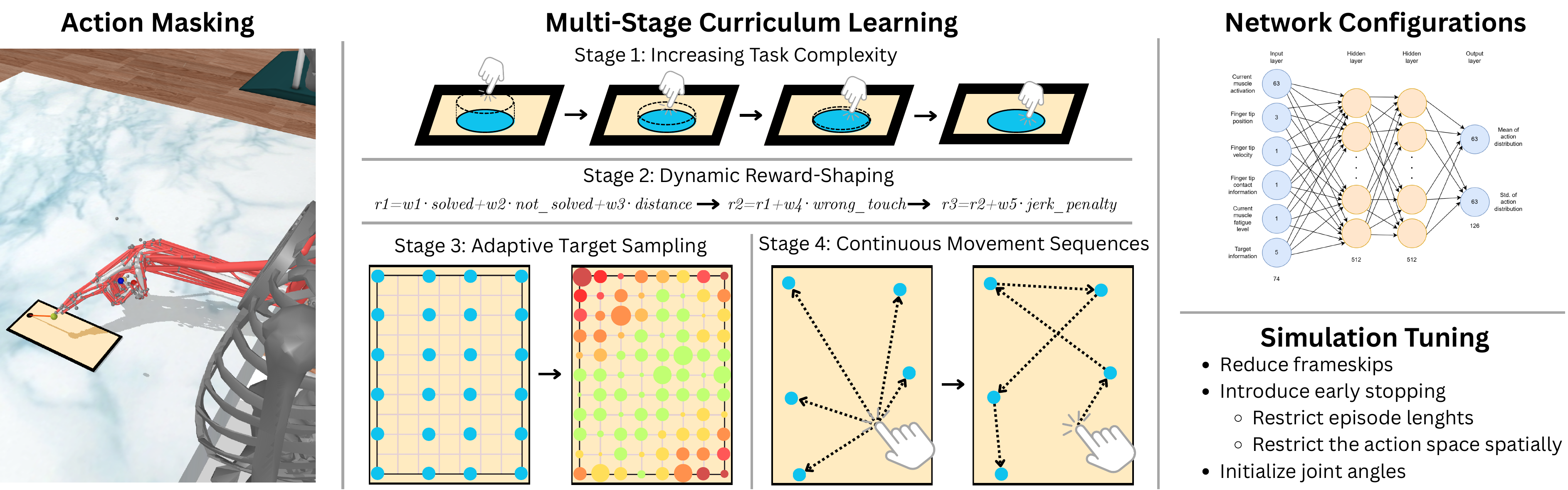}
  \caption{Overview of training routines to improve RL-based biomechanical forward simulations.}
  \label{fig:teaser}
\end{teaserfigure}

\maketitle

\section{Introduction}
\begin{table*}[ht]
    \small
    \centering
    \caption{Evaluation results represent the mean values of 100 episodes, each with a maximum duration of 10 seconds.} 
    \resizebox{\linewidth}{!}{
    \begin{tabular}{lccccccccc}
        \hline
        \makecell{\textbf{Network} \\ \textbf{Size}} & 
        \makecell{\textbf{Max} \\ \textbf{Timesteps}} & 
        \makecell{\textbf{Action} \\ \textbf{Masking}} &\makecell{\textbf{Curriculum} \\ \textbf{Learning}} & \makecell{\textbf{Stage 2} \\ \textbf{Dynamic Reward}}& \makecell{\textbf{Stage 2} \\ \textbf{Early Reward}}& \makecell{\textbf{Button} \\ \textbf{Radius (mm)}} & 
        \makecell{\textbf{Success} \\ \textbf{Rate}} & \makecell{\textbf{Avg. Errors /} \\ \textbf{Success Episode}}  & \makecell{\textbf{Avg. Time /} \\ \textbf{Success Episode (s)}}  \\
        \hline
        128 $\times$ 128 & 200M
        & \ding{51} & \ding{51} & \ding{55}     & \ding{51} & 1.5     
        & 82\% &  176.08 & 3.39 \\
        
        256 $\times$ 256  &  200M 
        & \ding{51} & \ding{51}      & \ding{51} & \ding{55} & 1.5     
        & 99\% &  156.81 & 3.35\\ 

        256 $\times$ 256 & 800M 
        & \ding{55} & \ding{51}      & \ding{51} & \ding{55}  & 1.5
        & 5\% & 3.6 & 1.6\\
        
        512 $\times$ 512 & 200M 
        & \ding{51} & \ding{51}      & \ding{51} & \ding{55} & 1.5     
        & 100\% & 8.07 &1.40\\ 
        
        512 $\times$ 512 & 200M 
        & \ding{51} & \ding{55}      & \ding{55} & \ding{55} & 6    
        & 0\% & 0 & \ding{55}  \\
        
        512 $\times$ 512 & 200M 
        & \ding{51} & \ding{55}      & \ding{55} & \ding{51}  & 1.5    
        & 64\% & 179.56 & 3.62\\
        
        \hline
    \end{tabular}
    }
    \label{tab:nn_params}
\end{table*}

Simulated users that model human-like, dexterous interactions with technology hold great potential to make user-centered design more efficient, rigorous, and predictable~\cite{smith2022what, lorenz_computational_2024}. At the core of these systems are biomechanical models that capture anatomical and physiological detail. Advances in computational power have made it possible to use increasingly complex models, such as those in \textit{MyoSuite}~\cite{caggiano2022myosuite, wang2022myosim} and \textit{User-in-the-Box}~\cite{ikkala2022breathing, ikkala2020converting}, which can simulate muscle fatigue or neuromuscular diseases.

However, training \gls{RL} agents to learn muscle-level control in physically simulated environments remains a highly complex task.
Training is often brittle, slow, and difficult to reproduce.
This limits interaction fidelity and accordingly the broader adoption of biomechanical modeling in the \gls{HCI} community for technology design and evaluation.
For instance, in \textit{User-in-the-Box}~\cite{ikkala2022breathing, fischer2021reinforcement}, the pointing targets had a radius of 5--15\,cm. Given that the iPhone 16 has a diagonal of 15.5\,cm this is large, especially when such models should be applied to mobile use cases.
Furthermore, the action space for arm movements was simplified by disabling wrist flexion and various muscles, effectively reducing the model to five degrees of freedom and 26 active muscles. 
In \textit{Sim2VR}, \citet{fischer2024sim2vr} also reduced the number of muscles and replaced some with joint torque actuators, which are easier to control. Their model also focuses primarily on coarse movements in VR.
Overall, current approaches demonstrate the potential of musculoskeletal forward simulation for \gls{HCI}, but they fall short of enabling the precise control required for most interactive tasks.
We argue that these shortcomings are primarily due to ineffective training routines.
Although reward shaping has been studied~\cite{selder2025makes, nowakowski2021human, chen2023static}, the influence of other training routines on the learning of fine motor skills remains underexplored.

In this work, we present and evaluate performance improvements achieved through action masking, curriculum learning, alternative policy network configurations, and task-specific improvements to the simulation environment. Based on our findings, we propose actionable training routines that lower technical barriers and support the development of high-fidelity, human-like control in physics-based biomechanical simulations for \gls{HCI}.

\section{Training Routines for Biomechanical Models}
Our primary goal is to learn stable muscle-level control policies that enable precise and dexterous movements. All evaluations are conducted using Proximal Policy Optimization (PPO)~\cite{schulman2017proximal}, and their performance is assessed based on the success rate of touch interactions. We consider an interaction as successful if the target element is touched within 10 seconds after the start of the episode.
We further count unintended interactions outside the target area (e.g., due to overshooting) as errors, indicating suboptimal control and a lack of fine motor skills. In the following, we introduce all the routines evaluated in this work.

\paragraph{\textbf{Action Masking}}
As shown in prior work~\cite{huang2020closer, miazga2025automated, stolz2024excluding, fischer2024sim2vr, ikkala2022breathing}, action masking is an effective technique to reduce the dimensionality of the action space, limit unnecessary exploration, and accelerate learning. In our implementation, we apply action masking by disabling all fingers except the index finger. Our results (see \autoref{tab:nn_params}) show that action masking highly improves training efficiency.

\paragraph{\textbf{Multi-Stage Curriculum Learning}} 
In \gls{RL}, curriculum learning improves learning efficiency and final performance by training agents on tasks of gradually increasing difficulty~\cite{narvekar2020curriculum, narvekar2017curriculum, luo2020accelerating}. To achieve the fine motor control required for touchscreen interactions, we introduce a structured, multi-stage curriculum. Each curriculum stage increases complexity via sub-stages.
The agent advances to the next (harder) stage only once it meets a performance threshold in the final sub-stage of the current curriculum stage. As shown in \autoref{tab:nn_params} and illustrated in \autoref{fig:teaser}, the following curriculum stages significantly improve training efficiency:

\textit{Stage 1: Increasing Task Complexity.} 
To simplify early training, the agent begins from a fixed position and must reach a large, button-shaped 3D target placed just above the surface with its fingertip.
As training progresses, the 3D target gradually flattens into a 2D surface (see \autoref{fig:teaser}), shifting the task from interacting with a volumetric object to a more demanding surface-based interaction that requires greater precision.

\textit{Stage 2: Dynamic Reward-Shaping.}
Reward shaping involves modifying the reward function to guide the agent toward desired behaviors more effectively~\cite{hu2020learning,grzes2017reward, vamplew_scalar_2022, dewey2014reinforcement, eschmann2021reward}. Our experiments show that starting with a simple reward to then gradually increase reward complexity, based on the training progress, significantly improves performance. Early rewards penalize incorrect button presses, while later stages discourage jerky movements and excessive muscle effort. Without this step-wise shaping, the agent fails to learn meaningful behavior.

\textit{Stage 3: Adaptive Target Sampling.}
Once the agent achieves satisfactory performance in the early stages of the curriculum, adaptive target sampling is used to improve generalization. The interaction surface is first discretized, and target sizes are randomized by uniformly sampling radii between 1.5\,mm and 7\,mm—encouraging the agent to learn the speed-accuracy trade-off observed in human interaction~\cite{wobbrock2008}. We also found that success rates vary significantly across target locations. To address this, we sample low-performing target locations more frequently, promoting balanced performance across the task environment.

\textit{Stage 4: Continuous Movement Sequences.}  
Human interaction with mobile devices such as smartphones can be seen as a sequence of movement primitives. Thus the agent must learn to perform any interaction movement from any starting position. To achieve this, each new target is sampled relative to the agent's current position and velocity.

\paragraph{\textbf{Network Configurations}}
Most current approaches~\cite{fischer2024sim2vr, ikkala2022breathing} are limited by the use of small policy networks, which struggle to capture the complexity of fine-motor tasks. While smaller networks are easier to train, they lack the capacity for precise control. In contrast, our training routines, combined with tuned optimizer settings, enable the use of significantly larger networks, leading to improved performance.
We set the learning rate and clip range to \(6 \times 10^{-4}\) and 0.2, respectively, and reduce both linearly over time. At timestep \(t\), each parameter $p$ is computed as: $p(t) = p_{t=0} \times r(t)$, where $r(t)$ linearly decays from 1 to 0 as training progresses. This schedule encourages exploration early on and gradually shifts toward more stable convergence.

Our results (\autoref{tab:nn_params}) further show that a larger policy network with a 512 × 512 architecture significantly outperforms smaller networks, enabling more accurate and stable fine-motor control.

\paragraph{\textbf{Simulation Tuning}}
We propose multiple small tweaks to the simulation environment, such as (1) reducing the frameskip parameter, which controls how frequently the agent receives observations and issues actions in MuJoCo~\cite{todorov2012mujoco, braylan2000frame}. We found that a low value (e.g., 3) allows fine-grained control, while higher values (e.g., 10) cause overshooting due to prolonged muscle activations. (2) We introduce early stopping criteria that terminate episodes whenever the agent exceeds the episode length restriction or leaves the permitted 3D action space constraint, thereby enforcing realistic temporal and spatial boundaries. (3) We initialize joint positions within the valid 3D action space, setting the index finger to point forward, to ensure consistent, task-relevant initial conditions.

\section{Conclusion}
This work demonstrates that tailored training routines and hyperparameter settings are key to achieving accurate and dexterous control in \gls{RL}-based musculoskeletal simulations. By evaluating task success and errors, we show that reliable fine-motor behavior is achievable, overcoming limitations of prior approaches. Our methods improve performance while lowering technical barriers, making biomechanical simulations more accessible for HCI research and applications. Implementations of our routines are available at: \url{https://github.com/ciao-group/RL-training-routines-for-Biomechanical-Models}.

\begin{acks}
The authors acknowledge the financial support by the Federal Ministry of Research, Technology and Space of Germany and by Sächsische Staatsministerium für Wissenschaft, Kultur und Tourismus in the programme Center of Excellence for AI-research „Center for Scalable Data Analytics and Artificial Intelligence Dresden/Leipzig“, project identification number: ScaDS.AI
\end{acks}


\balance
\bibliographystyle{ACM-Reference-Format}
\bibliography{references.bib}


\end{document}